\documentclass[aps,prl,reprint,preprintnumbers,superscriptaddress,amsmath,amssymb,bibnotes,longbibliography]{revtex4-1}

\usepackage{graphicx}% Include figure files
\usepackage{bm}% bold math
\usepackage[colorlinks,linkcolor=blue,anchorcolor=blue,citecolor=blue,urlcolor=blue,filecolor=blue,menucolor=blue,runcolor=blue]{hyperref}% add hypertext capabilities
\graphicspath{{../}}

\begin{document}

\preprint{Draft B. Shen resubmit}

\title{Pressure-induced dimerization and collapse of antiferromagnetism in the Kitaev material $\alpha$-Li$_2$IrO$_3$}% Force line breaks with \\

\author{Bin Shen}
\email{bin.shen@uni-a.de}
\affiliation  {Experimental Physics VI, Center for Electronic Correlations and Magnetism, University of Augsburg, 86159 Augsburg, Germany}

\author{Franziska Breitner}
\affiliation  {Experimental Physics VI, Center for Electronic Correlations and Magnetism, University of Augsburg, 86159 Augsburg, Germany}

\author{Danil Prishchenko}
\affiliation{Theoretical Physics and Applied Mathematics Department, Ural Federal University, 620002 Ekaterinburg, Russia}

\author{Rudra Sekhar Manna}
\affiliation  {Experimental Physics VI, Center for Electronic Correlations and Magnetism, University of Augsburg, 86159 Augsburg, Germany}
\affiliation  {Department of Physics, IIT Tirupati, Tirupati 517506, India}

\author{Anton Jesche}
\affiliation  {Experimental Physics VI, Center for Electronic Correlations and Magnetism, University of Augsburg, 86159 Augsburg, Germany}

\author{Maximilian L. Seidler}
\affiliation  {Experimental Physics VI, Center for Electronic Correlations and Magnetism, University of Augsburg, 86159 Augsburg, Germany}

\author{Philipp Gegenwart}
\email{philipp.gegenwart@uni-a.de}
\affiliation{Experimental Physics VI, Center for Electronic Correlations and Magnetism, University of Augsburg, 86159 Augsburg, Germany}

\author{Alexander A. Tsirlin}
\email{altsirlin@gmail.com}
\affiliation{Experimental Physics VI, Center for Electronic Correlations and Magnetism, University of Augsburg, 86159 Augsburg, Germany}

\date{\today}% It is always \today, today,

\begin{abstract}
We present magnetization measurements carried out on polycrystalline and single-crystalline samples of $\alpha$-Li$_2$IrO$_3$ under hydrostatic pressures up to 2~GPa and establish the temperature-pressure phase diagram of this material. The N\'eel temperature ($T_{\rm{N}}$) of $\alpha$-Li$_2$IrO$_3$ is slightly enhanced upon compression with $dT_{\rm{N}}/dp$ = 1.5~K/GPa. Above 1.2~GPa, $\alpha$-Li$_2$IrO$_3$ undergoes a first-order phase transition toward a nonmagnetic dimerized phase, with no traces of the magnetic phase observed above 1.8~GPa at low temperatures. The critical pressure of the structural dimerization is strongly temperature-dependent. This temperature dependence is well reproduced on the \textit{ab initio} level by taking into account lower phonon entropy in the nonmagnetic phase. We further show that the initial increase in $T_{\rm{N}}$ of the magnetic phase is due to a weakening of the Kitaev interaction $K$ along with the enhancement of the Heisenberg term $J$ and off-diagonal anisotropy $\Gamma$. Our study reveals a common thread in the interplay of magnetism and dimerization in pressured Kitaev materials.  
\end{abstract}

%pacs{Valid PACS appear here}% PACS, the Physics and Astronomy
\maketitle

%\tableofcontents

\section{Introduction}

The search for a quantum spin liquid (QSL), an exotic state characterized by long-range entanglement and fractionalized excitations, has been epitomized by the proposal of the honeycomb Kitaev model that offers an exact analytical solution for a QSL~\cite{06Kitaev}. In solid-state materials, this model can be realized in spin-orbit-coupled oxides and halides with the edge-sharing geometry of transition-metal octahedra~\cite{09JackeliPRL,19JackeliNRP}. The notable examples are Na$_2$IrO$_3$~\cite{10SinghPRB}, different polymorphs of Li$_2$IrO$_3$~\cite{tsirlin2021}, and $\alpha$-RuCl$_3$~\cite{21Loidl}. However, due to other interactions (such as Heisenberg $J$ and off-diagonal $\Gamma$ interactions) beyond the Kitaev term $K$~\cite{winter2017}, the aforementioned Kitaev candidates all display long-range magnetic order at low temperatures. The suppression of non-Kitaev interactions by a suitable tuning parameter is one of the possible strategies to reach the Kitaev limit and QSL.

At ambient pressure, $\alpha$-Li$_2$IrO$_3$ reveals the N\'eel temperature $T_{\rm{N}}$ = 15~K~\cite{12SinghPRL,17Freund,17MehlawatPRB} and develops incommensurate magnetic order with counter-rotating spin spirals~\cite{16WilliamsPRB}. Whereas low-energy excitations of this material resemble magnons~\cite{19ChoiPRB}, broader spectral features at higher energies were argued to arise from fractionalized excitations~\cite{revelli2020,20LiPRB} and may witness proximity to the Kitaev spin liquid, which might then be reached by a suitable tuning. Chemically tuned compounds, such as Ag$_3$LiIr$_2$O$_6$~\cite{19BahramiPRL} and H$_3$LiIr$_2$O$_6$~\cite{kitagawa2018} prepared by the ion exchange, feature disordered magnetic states indeed, but structural randomness~\cite{geirhos2020,bahrami2021} appears to be integral to the breakdown of magnetic order in these materials~\cite{tsirlin2021}. 

Hydrostatic pressure is a cleaner tuning parameter that does not introduce randomness but potentially drives structural phase transitions that necessarily affect magnetism. In this context, x-ray diffraction (XRD)~\cite{18HermannPRB, clancy2018, 20Layek}, optical spectroscopy~\cite{19HermannPRB}, and Raman spectroscopy~\cite{20LiPRB} on $\alpha$-Li$_2$IrO$_3$ defined $p_{\rm c}\simeq 3.8$~GPa as the critical pressure of the structural phase transition at room temperature. Above $p_{\rm c}$, $\alpha$-Li$_2$IrO$_3$ changes its symmetry from monoclinic to triclinic~\cite{18HermannPRB} and becomes non-magnetic owing to the formation of short Ir--Ir bonds (dimerization). This would leave a relatively broad window of nearly 4~GPa for tuning $\alpha$-Li$_2$IrO$_3$ without drastically changing its crystal structure.

Being common to honeycomb iridates, structural dimerization can lead to a complex behavior as a function of pressure and temperature. For example, in $\beta$-Li$_2$IrO$_3$ the dimerization also sets in at around 4~GPa at room temperature~\cite{19TakayamaPRB,choi2020}, but at low temperatures signatures of dimerization appear as low as 1.4~GPa~\cite{19VeigaPRB}, yet this dimerization is incomplete and results in a partially dimerized phase where only half of the Ir$^{4+}$ sites remain magnetic~\cite{19VeigaPRB,21Bin}. This intermediate phase mimics a pressure-induced spin liquid~\cite{18MajumderPRL}, but features cluster magnetism of decoupled spin tetramers rather than a collective entangled state of the Kitaev model~\cite{21Bin}.

Here, we investigate pressure-dependent magnetism of $\alpha$-Li$_2$IrO$_3$ at low temperatures. We find that $T_{\rm N}$ of \mbox{$\alpha$-Li$_2$IrO$_3$} increases with pressure and ascribe this effect to the enhancement of $J$ and $\Gamma$ along with the reduction in $K$ as the Ir--O--Ir bridging angles become smaller upon compression. Compared to the room-temperature data, the dimerization transition shifts to lower pressures as temperature is decreased. This shift is well reproduced on the \textit{ab initio} level and interpreted as the effect of reduced vibrational entropy caused by the hardening of phonons in the dimerized phase. We thus establish main trends in the pressure evolution of honeycomb iridates within both magnetic (nondimerized) and nonmagnetic (dimerized) phases.

%-------------------------------------------------------------------------------------------------------         
\section{Methods}

The presence or absence of the magnetic ordering transition in honeycomb iridates may depend on the sample quality~\cite{bahrami2021}. Therefore, we performed measurements on both polycrystalline and single-crystalline samples of $\alpha$-Li$_2$IrO$_3$. Previous work showed that single crystals contain the lowest amount of stacking faults and display a sharp transition at $T_{\rm N}$~\cite{17Freund}. This transition becomes broad in polycrystalline samples as the amount of stacking faults increases~\cite{Mannithesis}.

The polycrystalline sample of $\alpha$-Li$_2$IrO$_3$ was prepared by a solid-state reaction similar to Refs.~\cite{12SinghPRL,19ChoiPRB}. Single crystals of $\alpha$-Li$_2$IrO$_3$ were grown by the vapor transport method, as described in Ref.~\cite{17Freund}. Both polycrystalline sample and single crystals were characterized by XRD and ambient-pressure magnetization measurements, in which foreign phases, especially $\beta$-Li$_2$IrO$_3$, were not detected. 

The protocol for measuring magnetization under pressure was similar to the one implemented in Ref.~\cite{21Bin}. For measurements performed on single crystals, about ten randomly-oriented small crystals were inserted into the cell. Pressure was determined by measuring the superconducting transition of a small piece of Pb. Daphne oil was used as a pressure transmitting medium. 

Thermal expansion was measured at ambient pressure in the physical property measurement system (PPMS) using capacitive dilatometry with high resolution of 0.05~\r A at low temperatures~\cite{Kuchler2012}. The linear thermal expansion coefficient $\alpha$ = $d[\Delta L/L_0]/dT$ was determined from the differential length change. Measurement was carried out on a pellet pressed inside the glove box in order to avoid air trapping inside the pellet. Thermal expansion data were obtained upon warming with a temperature sweep rate of +0.3~K/min. 

On the \textit{ab initio} level, the structural phase transition in $\alpha$-Li$_2$IrO$_3$ was studied by full-relativistic density-functional (DFT) band-structure calculations performed in the \texttt{VASP}~\cite{vasp1,vasp2} code with the Perdew-Burke-Ernzerhof (PBE) for solids (PBEsol) exchange-correlation potential~\cite{pbesol} that allows the best agreement with the unit cell volume of $\alpha$-Li$_2$IrO$_3$ at ambient pressure. Correlation effects were taken into account on the DFT + $U$ + SO level with the on-site Coulomb repulsion parameter $U_d=1.0$~eV and Hund's coupling $J_d=0.3$~eV~\cite{21Bin}. Phonon spectra and corresponding thermodynamic functions were calculated in \texttt{Phonopy}~\cite{phonopy} using $2\times 2\times 2$ supercells with 0.01~\r{A} displacements. The $8\times 8\times 8$ and $4\times 4\times 4$ $k$-meshes were used for the atomic relaxations and phonon calculations, respectively.

An initial spin polarization with the ferromagnetic spin alignment was introduced for both dimerized and nondimerized phases. Calculations for the dimerized phase always converged to a nonmagnetic solution. On the other hand, magnetism could be stabilized in the nondimerized phase if a finite $U$ was applied, whereas calculations with $U=0$ produced a nonmagnetic solution that consequently evolved toward a dimerized structure when atomic positions and lattice parameters were optimized. A similar behavior has been previously reported in $\alpha$-RuCl$_3$ that also requires a finite $U$ to stabilize the magnetic phase~\cite{widmann2019}. Experimentally, $\alpha$-Li$_2$IrO$_3$ displays antiferromagnetic order at ambient pressure, but this incommensurate and noncoplanar magnetic structure~\cite{16WilliamsPRB} can not be incorporated in DFT. The ferromagnetic state is proximate to the incommensurate magnetic phase~\cite{baez2019,winter2016} and serves as a reasonable approximation when calculating thermodynamics of $\alpha$-Li$_2$IrO$_3$.

Additionally, we performed scalar-relativistic \texttt{FPLO} calculations~\cite{fplo} on the PBE level to obtain tight-binding parameters via Wannier projections. These calculations were run with fixed atomic positions determined from the structural optimizations in \texttt{VASP}. The denser $12\times 12\times 12$ $k$-mesh was used to ensure convergence.

\section{Experimental results}

 \begin{figure}
	\includegraphics[angle=0,width=0.48\textwidth]{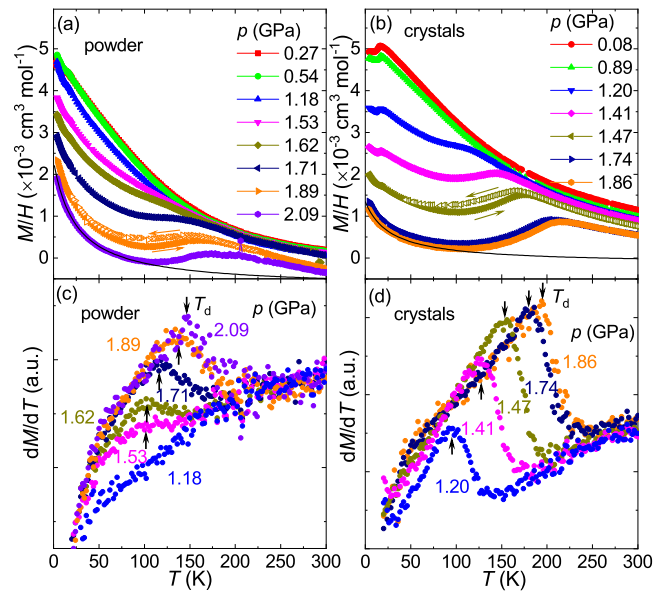}
		\vspace{-12pt} \caption{\label{Figure1} (a, b) Temperature-dependent dc magnetic susceptibility $M/H(T)$ of $\alpha$-Li$_2$IrO$_3$ measured under various pressures from 4 to 300~K in the 7-T magnetic field. Solid lines show the Curie-Weiss fits at low temperatures. (c, d) Temperature derivative of $M(T)$. The arrows denote the dimerization temperature $T_{\rm{d}}$. }
	\vspace{-12pt}
\end{figure}

Figure~\ref{Figure1} shows the dc magnetic susceptibility $M/H$ as a function of the temperature under various pressures for both samples. At low pressures, the single-crystalline sample displays a clear peak at $T_{\rm N}$. In contrast, the polycrystalline sample reveals only a broad shoulder, whereas magnetic susceptibility increases also below $T_{\rm N}$, likely due to defects.

At 1.53~GPa (1.20~GPa) for the polycrystalline (single-crystalline) sample, a peak appears around 100~K and shifts to higher temperatures upon further compression. This peak is a broadened step-like anomaly, which would be expected upon an abrupt dimerization~\cite{18BastienPRB} that renders Ir$^{4+}$ ions nonmagnetic. A hysteresis loop in $M/H(T)$ can be detected upon cooling and warming, indicating first-order nature of the transition and confirming its structural origin. Compared to $M/H(T)$ collected on powders, the data measured on single crystals show a sharper feature upon dimerization, owing to the better structural integrity of the single-crystalline sample. The sharper dimerization transition in single crystals can be also seen from d$M$/d$T$ [Figs.~\ref{Figure1}(c) and \ref{Figure1}(d)], where peak position is taken as the transition temperature $T_{\rm d}$.

At 1.2--1.8~GPa, both the dimerization transition and the low-temperature anomaly at $T_{\rm{N}}$ are observed, whereas above 1.8~GPa the anomaly at $T_N$ disappears. At the highest pressure reached in each of the runs, the low-temperature susceptibility is featureless and follows the Curie-Weiss law [Figs.~\ref{Figure1}(a) and \ref{Figure1}(b)], ${\chi=\chi_0+C/(T-\theta_{\rm CW})}$ where $\chi_0$ stands for the residual temperature-independent term, $C$ is the Curie constant, and $\theta_{\rm CW}$ is the Curie-Weiss temperature. The Curie-Weiss fit returns the effective moment of about 0.8 (0.6)~$\mu_B$/f.u. and $\theta_{\rm CW}\simeq -24$ ($-28$)~K in the polycrystalline (single-crystalline) sample at 2.09 (1.86)~GPa. These values are nearly constant in the narrow pressure range between the completion of the structural phase transition (1.7-1.8~GPa) and the highest pressure of our measurement. At lower pressures, the presence of the magnetic transition strongly affects the results of the Curie-Weiss fitting.

From the aforementioned Curie-Weiss parameters, using the effective moment of 1.83 (1.80)~$\mu_B$/f.u. extracted between 200 and 300~K at ambient pressure, we estimate that about 17 (10)\% of the weakly coupled $j_{\rm eff}=\frac12$ moments should be responsible for the Curie-like upturn in the dimerized phase at low temperatures. A similar low-temperature contribution was observed in pressurized $\beta$-Li$_2$IrO$_3$ and assigned to impurity spins, either intrinsic or introduced upon the compression~\cite{21Bin}. The higher concentration of the impurity spins in the polycrystalline sample of $\alpha$-Li$_2$IrO$_3$ is consistent with the presence of the susceptibility upturn below $T_{\rm N}$ already at ambient pressure.

 \begin{figure}
	\includegraphics[angle=0,width=0.48\textwidth]{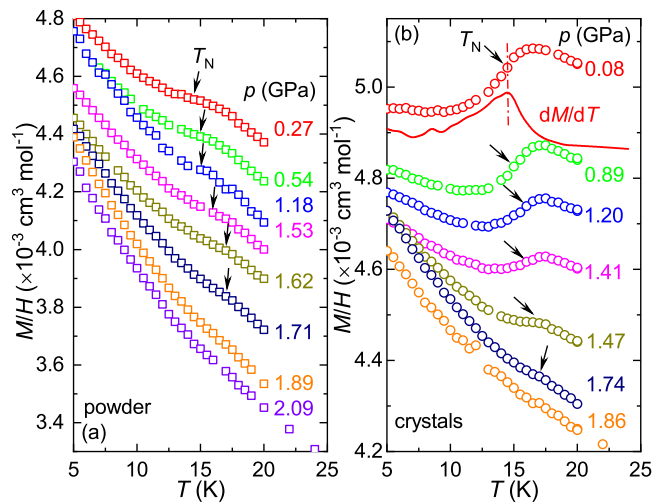}
	\vspace{-12pt} \caption{\label{Figure2} Low-temperature part of the $M/H(T)$ data for $\alpha$-Li$_2$IrO$_3$ measured on (a) polycrystalline and (b) single-crystalline samples. The arrows mark the antiferromagnetic transition temperature $T_{\rm{N}}$. The red solid line in  (b) is temperature derivative of $M(T)$ for the 0.08~GPa data. The data are vertically shifted for clarity.}
	\vspace{-12pt}
\end{figure}

Having established the pressure-induced dimerization in $\alpha$-Li$_2$IrO$_3$, we also address the evolution of $T_{\rm N}$ at lower pressures. The N\'eel temperature is determined from the peak position in $dM/dT$, as shown in Fig.~\ref{Figure2}. Even though the polycrystalline sample shows only a weak feature at $T_{\rm N}$, the magnetic ordering transition in this sample can be traced up to 1.71~GPa, similar to the single-crystalline sample. We find a good match between the $T_{\rm N}$ values in both samples and the systematic increase with the slope of 1.5~K/GPa. 

\begin{figure}
	\includegraphics[angle=0,width=0.4\textwidth]{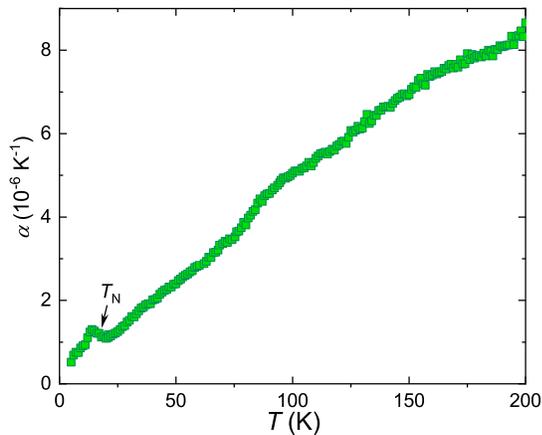}
	\vspace{-12pt} \caption{\label{Figure3} Thermal expansion coefficient of $\alpha$-Li$_2$IrO$_3$ measured as a function of temperature in zero field.}
	\vspace{-12pt}
\end{figure}

This pressure dependence of $T_{\rm N}$ can be cross-checked by thermal expansion measurement at ambient pressure. Figure~\ref{Figure3} shows the thermal expansion coefficient of $\alpha$-Li$_2$IrO$_3$. According to the Ehrenfest relation $dT_{\rm{N}}/dp$ = $V_{mol}$ $\times$ $T_{\rm{N}}$ $\times$ $\Delta \beta$ /  $\Delta C$ where $\Delta \beta$ and  $\Delta C$ are changes in, respectively, volume thermal expansion and specific heat upon the transition.  By using the molar volume $V_{mol}$ = 3.6 $\times$ 10$^{-5}$ m$^3$ mol$^{-1}$, $\Delta \beta$ = 3$\Delta \alpha$ = 1.8$\pm$1.0 $\times$ 10$^{-6}$~K$^{-1}$ (the uncertainty is derived from the broad transition in thermal expansion), and $\Delta C$ = 0.75~J mol$^{-1}$ K$^{-1}$ taken from Ref.~\cite{17Freund}, one can estimate a pressure dependence of the transition temperature of $\alpha$-Li$_2$IrO$_3$ in the zero pressure limit ($dT_{\rm{N}}/dp$)$_{p \rightarrow 0}$ = 1.3$\pm$0.7~K/GPa, which agrees well with our direct estimate of $T_{\rm{N}}(p)$ from magnetization measurements.

  \begin{figure}
 	\includegraphics[angle=0,width=0.48\textwidth]{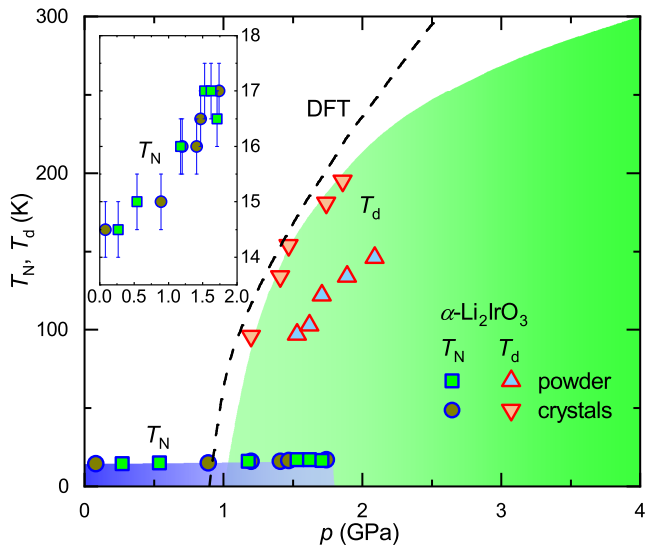}
 	\vspace{-12pt} \caption{\label{Figure4} Temperature-pressure phase diagram of $\alpha$-Li$_2$IrO$_3$ derived from the magnetic susceptibility data. $T_{\rm{N}}$ stands for the antiferromagnetic transition temperature, and $T_{\rm{d}}$ marks the dimerization transition temperature. The black dashed line marks the \textit{ab initio} estimate of $T_{\rm{d}}$ as a function of the pressure. The inset shows the pressure evolution of $T_{\rm{N}}$.}
 	\vspace{-12pt}
 \end{figure}

Based on the magnetization measurements, we construct the temperature-pressure phase diagram of $\alpha$-Li$_2$IrO$_3$, as depicted in Fig.~\ref{Figure4}. The region of the antiferromagnetically ordered phase increases upon compression following the increase in $T_{\rm N}$. Above 1.2~GPa, the nondimerized phase coexists with the dimerized phase, which is characterized by a phase boundary with the rapidly increasing $T_{\rm d}$. Above 1.8~GPa, the nondimerized phase disappears. Both single-crystalline and polycrystalline samples show a similar trend in $T_{\rm d}$, but the corresponding phase boundaries are shifted by 0.3-0.4~GPa relative to each other. This discrepancy may be caused by a broader dimerization transition in the polycrystalline sample and by the lower accuracy of $T_{\rm d}$ determined therein. 

Phase coexistence between 1.2 and 1.8~GPa is typical for a first-order phase transition. However, this hysteresis is notably broader than that in $\beta$-Li$_2$IrO$_3$ where a mixture of the magnetically ordered and partially dimerized phases was seen over a narrow pressure range of 0.2~GPa only~\cite{21Bin}. The expansion of the coexistence region in $\alpha$-Li$_2$IrO$_3$ may indicate a more substantial structural transformation upon the transition. Indeed, $\beta$-Li$_2$IrO$_3$ develops only a partially dimerized phase above $p_{\rm c}$~\cite{19VeigaPRB}, whereas in $\alpha$-Li$_2$IrO$_3$ one should expect a complete dimerization, as further confirmed by our \textit{ab initio} results below.

\section{\textit{Ab initio} modeling}

\subsection{Thermodynamic stability}
\begin{figure}
	\includegraphics{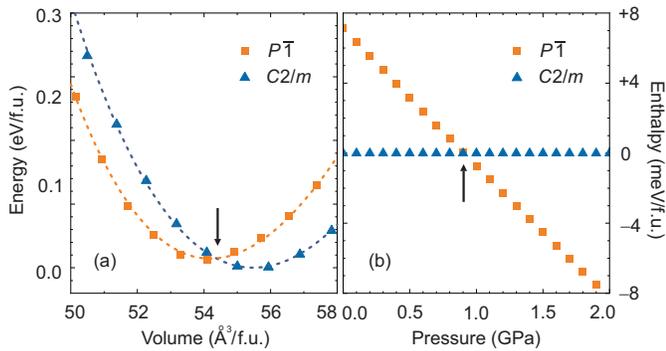}
	\caption{\label{fig:evsv}
		Energetics of the nondimerized ($C2/m$) and dimerized ($P\bar 1$) phases of $\alpha$-Li$_2$IrO$_3$: (a) Volume dependence of energy (symbols) and its fit with Eq.~\eqref{eq:eos} (dashed lines). (b) Pressure dependence of enthalpy. The arrows indicate the transition between the two phases. The nondimerized phase is chosen as reference.
	}
\end{figure}

In the following, we compare total energies of the magnetic nondimerized (monoclinic, $C2/m$) and nonmagnetic dimerized (triclinic, $P\bar 1$) phases of $\alpha$-Li$_2$IrO$_3$~\cite{18HermannPRB}. Figure~\ref{fig:evsv}(a) shows total energies calculated for fixed unit-cell volumes upon a full relaxation of the lattice parameters and atomic positions. The global energy minimum is that of the nondimerized phase, but a transformation toward the dimerized phase is expected upon compression. To assess the transition pressure, we fit energy-vs-volume curves with the Murnaghan equation of state,
\begin{align}
	E(V)=E_0+B_0V_0 &\left[\frac{1}{B_0'(B_0'-1)}\right. \left(\frac{V}{V_0}\right)^{1-B_0'}+ \notag\smallskip \\
	&\left.+\frac{1}{B_0'}\frac{V}{V_0}-\frac{1}{B_0'-1}\right].
	\label{eq:eos}\end{align}
where $E_0$ stands for the energy minimum for a given polymorph, $V_0$ is its equilibrium volume, $B_0$ is the bulk modulus, and $B_0'$ is the pressure derivative of $B_0$. The fitted parameters listed in Table~\ref{tab:eos} reveal the anticipated compression and lattice hardening upon dimerization. 

Equation of state parameters for the nondimerized phase show an excellent agreement with the experimental values determined from room-temperature x-ray diffraction, $B_0^{\exp}=106(5)$\,GPa and $V_0^{\exp}=55.03(3)$\,\r A$^3$~\cite{18HermannPRB}. The parameters for the dimerized phase show a slightly less favorable agreement, $B_0^{\exp}=125(3)$\,GPa and $V_0^{\exp}=53.6(2)$\,\r A$^3$~\cite{18HermannPRB}, probably because this phase could only be accessed above 4~GPa in the experiment, and its equilibrium volume could not be measured directly. Nevertheless, the trend of the lattice hardening is quite robust and well reproduced by our calculations. Moreover, we get the same dimerization pattern (all the $X$- or all the $Y$-bonds dimerized) and the shortest Ir--Ir distance of 2.63\,\r A at 4\,GPa in a reasonable agreement with the experimental value of 2.68(2)\,\r A~\cite{18HermannPRB}.

\begin{table}
	\caption{\label{tab:eos}
		Equation of state parameters for the nondimerized ($C2/m$) and dimerized ($P\bar 1$) phases of $\alpha$-Li$_2$IrO$_3$, see text and Eq.~\eqref{eq:eos}. The energies $E_0$ are given relative to the energy minimum of the $C2/m$ phase.
	}
	\begin{ruledtabular}
		\begin{tabular}{ccccc}
			Space group & $E_0$ (meV/f.u.) & $V_0$ (\r A$^3$/f.u.) & $B_0$ (GPa) & $B_0'$ \smallskip\\
			$C2/m$ & 0 & 55.48(2) & 103(1) & 4.5(3) \\
			$P\bar 1$ & 7(1) & 54.19(2) & 112(1) & 6.1(2) 
		\end{tabular}
	\end{ruledtabular}
\end{table}

Pressure-dependent enthalpy calculated from the parameters given in Table~\ref{tab:eos} reveals the dimerization transition at $p_c\simeq 0.9$~GPa [Fig.~\ref{fig:evsv}(b)]. This value of $\left.p_{\rm c}\right|_{T=0}$ corresponds to zero temperature, as no thermal effects have been taken into account. Such a critical pressure is much lower than 3.8~GPa expected from the previous room-temperature measurements~\cite{18HermannPRB,19HermannPRB,20LiPRB}, yet it nearly coincides with the low-temperature value of $p_c$ inferred from our magnetization data (Fig.~\ref{Figure4}). It is worth mentioning that DFT calculations for $\beta$-Li$_2$IrO$_3$ predicted a transformation from the nondimerized phase to the fully dimerized phase at around 2.2~GPa, and this critical pressure is clearly higher than 1.4~GPa determined experimentally for the same compound at low temperature~\cite{21Bin}. Such a discrepancy indicates that an intermediate partially dimerized phase should appear in $\beta$-Li$_2$IrO$_3$, and indeed this phase becomes stable around 1.4~GPa before giving way to the fully dimerized phase at higher pressures. On the other hand, in $\alpha$-Li$_2$IrO$_3$ the fully dimerized and nondimerized phases are much closer in energy, and no intermediate phase occurs.

Our experimental data further suggest the strong temperature dependence of $p_c$ (Fig.~\ref{Figure4}). We assess it by calculating thermodynamic functions for both phases of $\alpha$-Li$_2$IrO$_3$ in a harmonic approximation and adding both phonon energy and phonon entropy to the enthalpy difference shown in Fig.~\ref{fig:evsv}(b). Phonons were calculated for the 1~GPa crystal structures where enthalpies of both phases are nearly equal. 

\begin{figure}
	\includegraphics{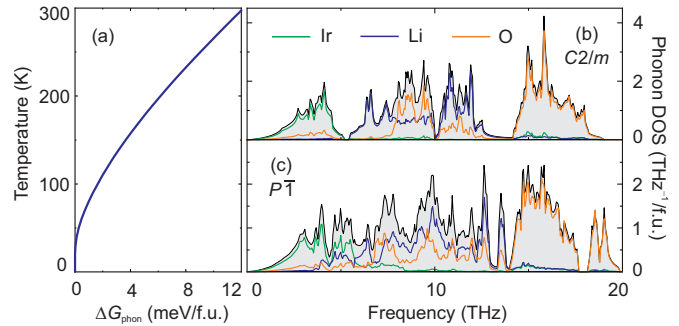}
	\caption{\label{fig:phonon}
		(a) Phonon free energy $\Delta G_{\rm phon}=G_{\rm phon}^{P\bar 1}-G_{\rm phon}^{C2/m}$. Positive values of $\Delta G_{\rm phon}$ indicate that at elevated temperatures phonon contribution to the free energy stabilizes the nondimerized phase. (b,c) Total and atomic-resolved phonon density of states for the (b) nondimerized and (c) dimerized phases.
	}
\end{figure}

Figure~\ref{fig:phonon}(a) shows that phonon free energy of the dimerized phase is systematically higher compared to the nondimerized phase. Their difference $\Delta G_{\rm phon}(T)=G_{\rm phon}^{P\bar 1}-G_{\rm phon}^{C2/m}$ increases with temperature, suggesting that difference in phonon entropy plays the main role. Even if the dimerized phase becomes stable at zero temperature above $\left.p_{\rm c}\right|_{T=0}=0.9$~GPa, at higher temperatures $\Delta G_{\rm phon}(T)>0$ renders the nondimerized phase more stable. This corresponds to the increase in $T_{\rm d}$ with pressure and, consequently, to the upward shift in $p_{\rm c}$ upon heating. 

\begin{table*}
	\caption{\label{tab:exchange}
		Pressure-induced changes in the ${X/Y}$-, ${Z}$-bonds distances $d$ (in \r A), Ir--O--Ir bridging angles $\varphi$ (in deg), and exchange parameters $J$, $K$, $\Gamma$ (in meV). 
	}
	\begin{ruledtabular}
		\begin{tabular}{cccccccccccc}
			$P$   & $d_{XY}$  & $\varphi_{XY}$ & $J_{XY}$ & $K_{XY}$ & $\Gamma_{XY}$ & & $d_{Z}$ & $\varphi_{Z}$ & $J_Z$  & $K_Z$  & $\Gamma_Z$  \smallskip\\
			0.0 GPa   & 2.988  &  94.89       & $-3.3$   & $-14.4$  &   11.2  &  & 2.968    &  93.57      & $-5.3$ & $-7.9$ &  14.2      \\
			0.8 GPa  & 2.980  &    94.67       & $-3.7$   & $-13.6$  &   11.8 &   & 2.958     &  93.28      & $-5.8$ & $-6.5$ &  15.0      \\
			difference  &            &     &            +12\%   &  $-6$\%  &   +5\%        &   &             &    &   +9\%  & $-18$\% & +6\%  \\
		\end{tabular}
	\end{ruledtabular}
\end{table*}

From the $\Delta G_{\rm phon}(T)$ values and the zero-temperature enthalpies at different pressures, we determine temperature dependence of $p_c$ that shows a remarkably good agreement with the experimental results for the single-crystalline sample (Fig.~\ref{Figure4}). At higher temperatures, the DFT prediction deviates from the experiment. Our calculation predicts the room-temperature dimerization at 2.6~GPa, at odds with the critical pressure of 3.8~GPa observed experimentally by x-ray diffraction~\cite{18HermannPRB}. This discrepancy may be caused by anharmonic effects that become more important at higher temperatures and also by the pressure dependence of the phonon energies neglected in our model. Nevertheless, even with this simple model we are able to pinpoint the origin of $G_{\rm phon}$ and of the temperature-dependent $p_c$. 

The difference in the phonon entropies should be traced back to the upward shift in the phonon energies upon dimerization [Figs.~\ref{fig:phonon}(b) and \ref{fig:phonon}(c)]. For the low-energy modes dominated by Ir, this shift is explained by the formation of the Ir--Ir bonds in the dimerized state. An equally strong shift seen for the Li- and O-related modes is caused by the overall deformation of the structure. For example, at 1~GPa the Ir--O distances are in the range of 2.02--2.03~\r A in the nondimerized phase and 2.00--2.05~\r A in the dimerized phase. Likewise, the spread of the Li--O distances increases from 2.07--2.15~\r A to 2.03--2.24~\r A, respectively. The shorter Ir--O and Li--O distances become possible in the dimerized phase and cause the hardening of the phonons. This effect correlates with the lattice hardening, as revealed by the increased bulk modulus ($B_0$) and its pressure derivative ($B_0'$) in the dimerized phase (Table~\ref{tab:eos}).

%---------------------------------------------------------------------------------------------------------
\subsection{Electronic structure}

\begin{figure}
\includegraphics{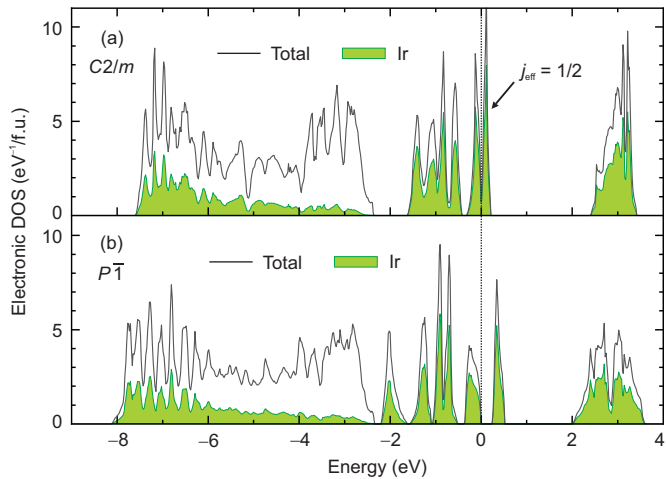}
\caption{\label{fig:DOS}
Electronic density of states calculated on the DFT+SO level for the fully optimized crystal structures of the (a) nondimerized phase at 0~GPa and (b) dimerized phase at 4~GPa. The Fermi level is at zero energy. Note that the nondimerized phase is metallic in DFT+SO and thus requires Hubbard $U$ as well as magnetism in order to open the band gap, whereas the dimerized phase is insulating already in DFT+SO owing to the formation of quasi-molecular orbitals on the Ir--Ir dimers.
}
\end{figure}

According to Refs.~\cite{clancy2018,19TakayamaPRB,antonov2018,antonov2021}, the dimerization leads to a major change in the electronic structure of Kitaev materials. Indeed, band structures of $\alpha$-Li$_2$IrO$_3$ calculated on the DFT+SO level using fully optimized crystal structures of the nondimerized phase at 0~GPa and the dimerized phase at 4~GPa reveal a major reconstruction of the Ir $t_{2g}$ states in the vicinity of the Fermi level (Fig.~\ref{fig:DOS}). In the absence of $U$, the nondimerized phase shows a metallic energy spectrum with a relatively narrow $j_{\rm eff}=\frac12$ band that can be split by a moderate $U$ to induce the insulating state observed experimentally. On the other hand, the dimerized phase is insulating even on the DFT+SO level, because the Ir $t_{2g}$ states transform into several narrow bands that manifest the formation of quasi-molecular orbitals of the Ir--Ir dimers.

%---------------------------------------------------------------------------------------------------------
\subsection{Magnetism}

We will now discuss why $T_N$ of the nondimerized phase increases with pressure. To this end, we analyze pressure-induced changes in the atomic positions and exchange interactions using the crystal structures obtained by a full relaxation at 0 and 0.8~GPa. Different microscopic models were proposed to explain magnetism of $\alpha$-Li$_2$IrO$_3$ and its unusual magnetic structure~\cite{chaloupka2015,kimchi2015,winter2016,nishimoto2016,baez2019}. While there is no consensus on the relevant interaction terms and the role of interactions beyond nearest neighbors, one expects that nearest-neighbor couplings mediated by the Ir--O--Ir bridges show a much stronger pressure dependence than any of the long-range couplings~\cite{kaib2021}. Therefore, pressure dependence of $T_{\rm N}$ should be mainly caused by pressure-induced changes in the parameters of the nearest-neighbor spin Hamiltonian
\begin{equation*}
	\mathcal H=\sum_{\langle ij\rangle} J_{ij}\mathbf S_i\mathbf S_j + \sum_{\langle ij\rangle} K_{ij} S_i^{\gamma}S_j^{\gamma}+\sum_{\langle ij\rangle} \Gamma_{ij} (S_i^{\alpha}S_j^{\beta}+S_i^{\beta}S_j^{\alpha}),
\end{equation*}
where $J_{ij}$, $K_{ij}$, and $\Gamma_{ij}$ stand, respectively, for the Heisenberg exchange, Kitaev exchange, and off-diagonal anisotropy, and $\alpha\neq\beta\neq\gamma$.

We utilize superexchange theory developed in Ref.~\cite{rau2014} and take advantage of its extension reported in Ref.~\cite{winter2016} in order to calculate the parameters $J$, $K$, and $\Gamma$ for the two nonequivalent nearest-neighbor bonds that correspond to the $X/Y$- and $Z$-bonds of the Kitaev model, respectively. A significant difference between these bonds is in agreement with the recent measurement of the magnetic diffuse scattering~\cite{chun2021} that also indicated a departure of spin-spin correlations from the three-fold symmetry and the sizable in-plane bond anisotropy in \mbox{$\alpha$-Li$_2$IrO$_3$}. We note in passing that the 0~GPa values in Table~\ref{tab:exchange}	are slightly different from those reported in Ref.~\cite{winter2016} because we did not include crystal-field terms and interactions due to multiple hoppings and, further, used the crystal structure relaxed at a constant pressure.

Table~\ref{tab:exchange} compares the Ir--O--Ir bridging angles as well as the $J$, $K$, $\Gamma$ parameters.  Lattice compression shortens the Ir--Ir distances and reduces the Ir--O--Ir angles. Whereas $J$ and $\Gamma$ increase in magnitude, the $K$ values decrease. These trends are well in line with theoretical expectations of the $|K|\ll \Gamma$ regime as the bridging angle approaches $90^{\circ}$~\cite{winter2016}. 

Considering that there are three $J$-terms and three $\Gamma$-terms per bond vs. one $K$-term per bond, we estimate, on average, the 17.5\% increase in the coupling energy on the $X$- and $Y$-bonds and the 7.5\% increase in the coupling energy on the $Z$-bonds. These changes are compatible with the 8\% increase in $T_N$ in the same pressure range. Therefore, we conclude that the increasing $T_N$ can be a result of the decreasing Ir--O--Ir bridging angles, which enhance the more abundant $J$ and $\Gamma$ interaction terms and reduce the less abundant $K$ terms in $\alpha$-Li$_2$IrO$_3$. 

\section{Discussion and Summary}

Our study elucidates two major aspects of the pressure evolution of honeycomb iridates. Their magnetic phase shows a systematic increase in $T_{\rm N}$ upon compression, with $dT_{\rm N}/dP$ of 0.7~K/GPa ($\beta$-Li$_2$IrO$_3$~\cite{18MajumderPRL}), 1.5~K/GPa ($\alpha$-Li$_2$IrO$_3$, present study), and 1.7~K/GPa (Na$_2$IrO$_3$~\cite{simutis2018}). Microscopically, this increasing $T_{\rm N}$ can be ascribed to the reduction in the Ir--O--Ir angles upon the shortening of the Ir--Ir bonds and the overall lattice contraction. Indeed, as the Ir--O--Ir angles decrease toward $90^{\circ}$, one expects the increase in the absolute values of $J$ and $\Gamma$~\cite{winter2016}. Even if $|K|$ concomitantly decreases, the overall coupling energy increases and boosts $T_{\rm N}$. Additionally, the reduction in $|K|$ implies that hydrostatic pressure tunes honeycomb iridates away from the Kitaev limit~\cite{tsirlin2021}.

At higher pressures, dimerization instability comes into play and drives a structural phase transition toward a nonmagnetic phase with the Ir--Ir dimers. An interesting and initially unexpected feature of this transition is the strong temperature dependence of $p_{\rm c}$ in both $\alpha$- and $\beta$-polymorphs of Li$_2$IrO$_3$. While the dimerization is observed around 4~GPa at room temperature, it starts already at 1.0--1.5~GPa when temperature is decreased. We explain this effect by a phonon contribution to the free energy. The phonons of the dimerized phase are harder, thus leading to a lower phonon entropy and a higher phonon free energy compared to the nondimerized phase. This difference causes the systematic upward shift of $p_{\rm c}$ upon heating. The same reasoning should apply to $\alpha$-RuCl$_3$ where $p_{\rm c}$ also shows a strong temperature dependence~\cite{18BastienPRB,cui2017}. Moreover, a similar phenomenology may be expected in Na$_2$IrO$_3$, albeit at much higher pressures, because larger unit-cell volume should shift the dimerization transition to pressures on the order of 50~GPa~\cite{18HuPRB}.

A comparison between the single-crystalline and polycrystalline samples of $\alpha$-Li$_2$IrO$_3$ reveals that neither pressure evolution of $T_{\rm N}$ nor the onset of the dimerized phase are affected by the stacking faults, although an increased concentration of these structural defects broadens all transition anomalies and renders them less discernible than in high-quality single crystals. Therefore, pressure-induced dimerization should be expected even in honeycomb iridates like Ag$_3$LiIr$_2$O$_6$ and H$_3$LiIr$_2$O$_6$ that are strongly affected by structural randomness. Indeed, recent pressure work on Cu$_2$IrO$_3$ revealed the formation of Ir--Ir dimers despite abundant stacking faults in that compound~\cite{fabbris2021}. 

In summary, we have shown that hydrostatic pressure drives $\alpha$-Li$_2$IrO$_3$ away from the Kitaev limit and at low temperatures causes a structural dimerization already at 1.2~GPa. No pressure-induced spin-liquid state occurs in this compound, similar to $\beta$-Li$_2$IrO$_3$. The dimerization transition is first-order in nature and evolves in the same way as in several other Kitaev candidates. This generic behavior is explained microscopically by the phonon hardening in the dimerized phase. An interesting question for future studies would be the fate of $\gamma$-Li$_2$IrO$_3$ upon compression. The suppression of magnetic Bragg peaks above 1.4~GPa without any change in the lattice symmetry~\cite{17BreznayPRB} would on one hand manifest a difference from the $\alpha$- and $\beta$-polymorphs where dimerization is accompanied by a symmetry lowering. On the other hand, the critical pressure of 1.4~GPa is conspicuously close to the zero-temperature critical pressures of the other two polymorphs, whereas structural dimerization in honeycomb iridates does not require a symmetry lowering~\cite{19VeigaPRB}. Magnetization measurements on $\gamma$-Li$_2$IrO$_3$ under pressure could shed further light on this issue.

\section{Acknowledgments}
This work was funded by the German Research Foundation (DFG) via Project No. 107745057 (TRR80) and via the Sino-German Cooperation Group on Emergent Correlated Matter. D.P. acknowledges financial support by the Russian Science Foundation, Grant No. 21-72-10136.

\end{document}